\documentclass[aps, notitlepage, pra, twocolumn, nofootinbib, floatfix, superscriptaddress, english, 10pt]{revtex4-1}
\usepackage{cancel}
\pdfoutput=1
\usepackage{booktabs}
\usepackage{graphicx,graphics,epsfig}   
\usepackage{dcolumn}    
\usepackage{bm}         
\usepackage{amsmath}    
\usepackage{verbatim}   
\usepackage{color}      
\usepackage{subfigure}  
\usepackage{amsmath,amsfonts,amssymb,graphics,graphics,color,times}

\usepackage[T1]{fontenc} 

\usepackage{xcolor} 
\usepackage{url}

\usepackage{latexsym}
\usepackage{amsmath}
\usepackage{amssymb}
\usepackage{amsfonts}
\usepackage{amsthm}
\usepackage{mathrsfs}
\usepackage{color,verbatim,graphics}
\usepackage{psfrag}
\DeclareMathAlphabet{\mathrsfs}{U}{rsfs}{m}{n}
\DeclareMathAlphabet{\mathpzc}{OT1}{pzc}{m}{it}
\DeclareMathAlphabet{\matheus}{U}{eus}{m}{n}
\DeclareMathAlphabet{\mathbbold}{U}{bbold}{m}{n}

\def\one{\leavevmode\hbox{\small1\normalsize\kern-.33em1}}

\newcommand{\RR}{\mathbb{R}}

\renewcommand{\qed}{\ensuremath{\hfill \blacksquare}}

\def\one{\leavevmode\hbox{\small1\normalsize\kern-.33em1}}

\newcommand{\ba}{\begin{eqnarray}}
\newcommand{\ea}{\end{eqnarray}}
\newcommand{\ban}{\begin{eqnarray*}}
\newcommand{\ean}{\end{eqnarray*}}
\newcommand{\Tr}{\operatorname{Tr}}

\newcommand{\ket}[1]{|#1\rangle}
\newcommand{\bra}[1]{\langle#1|}

\begin{document}

\title{Bounding the detection efficiency threshold in Bell tests using multiple copies of the maximally entangled two-qubit state carried by a single pair of particles}

\author{Istv\'an M\'arton}
\email{marton.istvan@atomki.hu}
\affiliation{MTA ATOMKI Lend\"ulet Quantum Correlations Research Group, Institute for Nuclear Research, P.~O. Box 51, H-4001 Debrecen, Hungary} 

\author{Erika Bene}
\email{bene@atomki.hu}
\affiliation{MTA ATOMKI Lend\"ulet Quantum Correlations Research Group, Institute for Nuclear Research, P.~O. Box 51, H-4001 Debrecen, Hungary} 

\author{Tam\'as V\'ertesi}
\email{tvertesi@atomki.hu}
\affiliation{MTA ATOMKI Lend\"ulet Quantum Correlations Research Group, Institute for Nuclear Research, P.~O. Box 51, H-4001 Debrecen, Hungary} 



\begin{abstract}

In this paper, we investigate the critical efficiency of detectors to observe Bell nonlocality using multiple copies of the maximally entangled two-qubit state carried by a single pair of particles, such as hyperentangled states, and the product of Pauli measurements. It is known that in a Clauser-Horne-Shimony-Holt (CHSH) Bell test the symmetric detection efficiency of $82.84\%$ can be tolerated for the two-qubit maximally entangled state. We beat this enigmatic threshold by entangling two particles with multiple degrees of freedom. The obtained upper bounds of the symmetric detection efficiency thresholds are $80.86\%$, $73.99\%$ and $69.29\%$ for two, three and four copies of the two-qubit maximally entangled state, respectively. The number of measurements and outcomes in the respective cases are 4, 8 and 16. To find the improved thresholds, we use large-scale convex optimization tools, which allows us to significantly go beyond state-of-the-art results.
The proof is exact up to three copies, while for four copies it is due to reliable numerical computations. Specifically, we used linear programming to obtain the two-copy threshold and the corresponding Bell inequality, and convex optimization based on Gilbert's algorithm for three and four copies of the two-qubit state. We show analytically that the symmetric detection efficiency threshold decays exponentially with the number of copies of the two-qubit state. Our techniques can also be applied to more general Bell nonlocality scenarios with more than two parties.  

\end{abstract}

\maketitle

\section{Introduction}\label{sec:intro}

Quantum theory predicts that there exist correlations in nature that cannot be simulated with classical resources. In particular, measurements on separated parts of entangled quantum systems can produce outcomes whose correlations cannot be explained by any local classical model. These strong correlations can be witnessed by violating Bell inequalities~\cite{Bell64}.

The violation of Bell inequalities was demonstrated in several laboratory experiments over the last 50 years. The very first conclusive experiment has been performed by Freedman and Clauser~\cite{FC} in 1972. However, this experiment did not close the locality loophole. Note that Bell experiments can have a number of loopholes due to technical imperfections~\cite{larsson14}. The first pioneering experiment with time-varying polarization analyzers -- still not loophole-free -- was carried out by Aspect and his colleagues~\cite{Aspect} in 1982. In fact, loophole-free Bell violations have only recently become possible~\cite{hensen15,giustina15,shalm15,rosenfeld2017}. These latter experiments provide the strongest evidence yet that nature is nonlocal.

One of the main technical difficulties in achieving a loophole-free violation is the finite detection efficiency of the detectors~\cite{pearle}. In a Bell test, due to the imperfect efficiency of the detectors, some of the emitted systems are not detected during the detection process, and this failure can be exploited by a local classical model to reproduce the statistics of the experiment. For each Bell inequality, a threshold can be computed above which the detection loophole closes (see for example the recent works in
Refs.~\cite{sciarrino11,kost18,cope19a}). For a typical Bell inequality, however, this threshold is relatively high. 
The first detection-loophole-free Bell tests~\cite{hensen15,giustina15,shalm15} performed in 2015 were all based on the violation of the Clauser-Horne-Shimony-Holt (CHSH) Bell inequality~\cite{CHSH}. In fact, this is the simplest bipartite Bell inequality, consisting of two settings ($m=2$) with two outcomes ($o=2$) per party, and the detection efficiency required to see a Bell violation is at least $\eta=2(\sqrt 2-1)\simeq0.8284$ using the maximally entangled two-qubit state~\cite{mermin1986,garg87}:
\begin{equation}
\ket{\varphi_+}=\frac{\ket{0,0}+\ket{1,1}}{\sqrt 2}.
\label{singlet}
\end{equation}
To determine the threshold $2(\sqrt 2-1)$, one assumes that all detectors have the same detection efficiency $\eta$. Using Bell inequalities with more than two settings per party ($m>2$) and a two-qubit maximally entangled state, only minor improvements were reported. To the best of our knowledge, the lowest reported symmetric detection efficiency threshold is $0.8214$ for two measurement outcomes and a maximally entangled two-qubit system~\cite{brunner_gisin}. This efficiency threshold is given by the four-setting Bell inequality $A_5$ from the list of Avis et al.~\cite{avis05}. Exactly the same threshold was obtained by Massar et al.~\cite{massar02} in the case of four settings, who used a geometric approach to study Bell inequalities with multiple settings. We also note that for the maximally entangled state~(\ref{singlet}), a lower bound of $2/3$ on the threshold $\eta$ has been proved for any number of two-outcome settings~\cite{gisin99}. However, it is not known whether this lower bound of $2/3$ is achievable, or whether a higher lower bound can be obtained using the maximally entangled state~(\ref{singlet}) and possibly an infinite number of measurement settings. 

In the present work, we study Bell inequalities with finite detection efficiencies using multiple copies of the two-qubit maximally entangled state~(\ref{singlet}), which state is carried by a single pair of particles. Note that $n$ copies of the state~(\ref{singlet}) (or any other two-qubit maximally entangled state) can be considered as a maximally entangled $(2^n\times 2^n)$-dimensional state. We use this $n$-copy state along with a product of Pauli measurements acting on the $n$ individual copies. The resulting probabilities define a Bell setup with $m=2^n$ settings and $o=2^n$ outcomes. Let us denote by $\eta_{sym}^{(n)}$ the lowest possible symmetric detection efficiency threshold in this $n$-copy scenario. By symmetric, we mean that each detector is modeled by the same detection efficiency. In this work, we give upper bounds on the above threshold. We can reduce the critical detection efficiency value well below $0.8214$ for $n\ge 2$. In particular, the critical value $\eta_{sym}^{(n)}$ is shown to decrease exponentially in the number of $n$ copies. To this end, in section~\ref{sec:multichsh}, we consider multiple copies of the CHSH expression. This inequality is sufficient to show the exponentially decreasing behavior of the critical value. 

In sections IV and V, however, we give even lower upper bounds on $\eta_{sym}^{(n)}$ by constructing optimal Bell inequalities based on a geometric approach. In particular, we obtain upper bounds of $\eta_{sym}^{(2)}\le 0.8086$ and $\eta_{sym}^{(3)}\le 0.7399$ for two and three copies, respectively. It is noted that for a single copy, we have the exact value $\eta_{sym}^{(1)} = 2(\sqrt 2-1)$, which is provided by the CHSH inequality. It is difficult to obtain exact bounds for more than three copies due to the computation of the local bound, but our reliable numerical calculations strongly support an upper bound of $\eta_{sym}^{(4)}\le 0.6929$. We also consider the case when the efficiency of one party's detectors is unity (i.e., say, Bob has perfect detectors), in which case we denote the corresponding $n$-copy threshold by $\eta_{asym}^{(n)}$. In this case, we also obtain improved upper bounds compared to the one-copy threshold $\eta_{asym}^{(1)} = 1/\sqrt 2 \simeq 0.7071$ (see, e.g., ~Refs.~\cite{detasym1,detasym2}). Note that our approach apart from the above cases is also applicable in the more general case when the detection efficiencies depend on the settings~\cite{garbarino10}. Such a more general setup fits for example to Bell experiments using hybrid measurements.   

In addition to the numerical treatment, we analytically upper bound $\eta_{sym}^{(n)}$ and find that the value $2/3$ in Gisin and Gisin's paper~\cite{gisin99} can be surpassed by $n\ge 13$ copies of the maximally entangled two-qubit state and $m=2^{13}$ measurement settings per party. This result also indicates how difficult it is to significantly reduce the required detection efficiency in the number of settings to observe a Bell violation in the $n$-copy scenario. In another high-dimensional scenario, Massar~\cite{massar02nonlocality} obtained the threshold values 
\begin{equation}
\eta_{sym}\simeq D^{3/4}2^{-0.0035D}
\end{equation}
for a special family of $2^D$-setting Bell inequalities using $(D\times D)$-dimensional maximally entangled states. If we consider this high-dimensional entangled state to be many copies of the maximally entangled two-qubit state~(\ref{singlet}), we find that $n$ must be greater than 10 (and the number of settings $m=2^D$ greater than the astronomically large number $2^{1024}$) to make $\eta_{sym}<2/3$.
 
Another approach to the detection efficiency problem, which is particularly fruitful in photonic experiments~\cite{giustina15,shalm15}, is to use partially entangled states instead of the maximally entangled two-qubit state. By modifying the state (and the measurements) in this way, the detection efficiency threshold for the CHSH inequality~\cite{CHSH} can be reduced from $0.8284$ down to $2/3$, the so-called Eberhard limit~\cite{eberhard93}. However, to reach this value, both the state and the applied measurements must be fine-tuned, which is experimentally challenging. Indeed, in the limiting case of $2/3$, the state has to be a pure state close to the product state and the measurements must correspond to observables which are almost commuting~\cite{eberhard93,hardy93}. Related to this, a recent experimental study has shown that almost product states are very fragile for obtaining a high rate of random bits~\cite{gomez19} from Bell violations~\cite{colbeck09,pironio10}.  

In this study, we take a somewhat opposite approach to the one using partially entangled states discussed above. We consider a measurement setup with products of Pauli observables on both sides and multiple copies of the two-qubit maximally entangled state~(\ref{singlet}) encoded into a pair of particles. It is noted that the state~(\ref{singlet}) is equivalent to the singlet state up to a local change of basis, which gives the threshold value of $\eta_{sym}=2(\sqrt 2 - 1)$ for the CHSH inequality~\cite{CHSH}. In the next section, we reproduce this known value~\cite{mermin1986,garg87}. We then show that this value can be reduced significantly by considering multiple copies of the maximally entangled two-qubit state and by performing anticommuting Pauli measurements on half of the two-qubit state. We note that the decreasing behavior of the detection efficiency threshold using multiple copies of the two-qubit state has been conjectured by Barrett et al.~\cite{barrett2002}.

Note that for two copies of the maximally entangled two-qubit state, i.e., a maximally entangled $4\times 4$ state, the lowest detection efficiency threshold found so far corresponds to the inequality $I_{4422}^{4}$~\cite{brunner_gisin}, which gives the threshold efficiency of $\eta_{sym}=0.7698$~\cite{VPN10}. 
In fact, more recent studies using Bell inequalities with multiple outcomes and higher dimensional maximally entangled states have found no improvement on this value~\cite{cope19a,cope19b}. While our results do not improve on previous works for the maximally entangled $4\times 4$ dimensional state, by adding a third copy of the maximally entangled two-qubit state to obtain an $8\times 8$ state encoded in a single pair of particles, we achieve $\eta_{sym}^{(3)}\le 0.7399$. Our result is promising from an experimental point of view, considering recent progress on hyperentanglement~\cite{kwiat97,erhard2020} and various sources of high-dimensional entanglement (see Ref.~\cite{genovese08} for a review).

The structure of the paper is as follows. In section~\ref{sec:belldet}, we introduce two-party correlation-type Bell inequalities (and in particular the CHSH inequality) and discuss their detection efficiency threshold. In section~\ref{sec:multichsh}, we present an $n$-th iterated version of the CHSH inequality and give upper bounds on the detection efficiency thresholds of this particular inequality. This in turn defines an upper bound on $\eta_{sym}^{(n)}$ and $\eta_{asym}^{(n)}$. In Secs.~\ref{sec:geom} and \ref{sec:geomgilbert} we present our geometric approaches based on convex optimization (linear programming and Gilbert's method). These methods allow us to further improve the above upper bounds for small values of $n$ (e.g. $n=2,3$ and $4$). The paper concludes with a discussion in Sec.~\ref{sec:discuss}. 

\section{Bell inequalities and detection efficiencies}\label{sec:belldet}

{\it The CHSH inequality.}---We use the maximally entangled two-qubit state~(\ref{singlet}), locally equivalent to the singlet state $(\ket{0,1}-\ket{1,0})/\sqrt 2$. On the other hand, Alice's and Bob's measurements are
\begin{align}
M_{a|x}&=(\one+(-1)^a\vec a_x\cdot\vec\sigma)/2,\nonumber\\
M_{b|y}&=(\one+(-1)^b\vec b_y\cdot\vec\sigma)/2, 
\label{anticomm}
\end{align}
where the outputs are labeled by $a,b=\{0,1\}$ and the inputs by $x,y=\{0,1\}$, and $\vec\sigma=(\sigma_x,\sigma_y,\sigma_z)$ is the vector of Pauli matrices. With the Bell state~(\ref{singlet}) and measurement directions (i.e. the Bloch vectors) $\vec a_x$ for Alice and $\vec b_y$ for Bob, the correlations are given by
\begin{align}
P(a,b|x,y)&=\Tr{\left(\ket{\varphi_+}\bra{\varphi_+}M_{a|x}\otimes M_{b|y}\right)}\nonumber\\
&=\frac{1}{4}\left(1+(-1)^{a\oplus b}\vec a_x\cdot\vec b'_y\right), 
\label{Pabxy}
\end{align}
where $\vec{b'}_y=(b_1^y,-b_2^y,b_3^y)$ and $a \oplus b$ is the sum of bits $a$ and $b$ modulo 2.

If we choose the measurement directions 
\begin{align}
\vec a_0&=(1,0,0)\nonumber\\
\vec a_1&=(0,0,1)
\label{blochA}
\end{align}
on Alice's side and measurement directions
\begin{align}
\vec b_0&=(1,0,1)/\sqrt 2\nonumber\\
\vec b_1&=(1,0,-1)/\sqrt 2
\label{blochB}
\end{align}
on Bob's side, we obtain the following statistics
\begin{equation}
P(a,b|x,y)=\frac{1}{4}\left(1+(-1)^{a\oplus b}(-1)^{xy}\frac{\sqrt 2}{2}\right),
\label{Pchsh}
\end{equation}
where $a,b,x,y$ are assumed to have values in $\{0,1\}$. These correlations give the symmetric detection efficiency threshold $\eta_{sym}=2(\sqrt 2 - 1)$~\cite{mermin1986,garg87}. To derive this value, let us first consider the more general case where Alice detects her particle with an efficiency $\eta_A$ and Bob detects his particle with an efficiency $\eta_B$ for all their input settings. In the special symmetric case $\eta_{sym}=\eta_A=\eta_B$. 

Let us write the CHSH inequality~\cite{CHSH} in the form~\cite{CGLMP02}: 
\begin{align}
{\rm CHSH}(a,b,x,y) =& P(00|00)+P(11|00) + P(00|01)\nonumber\\
&+P(11|01) + P(00|10)+P(11|10)\nonumber\\
&+ P(01|11)+P(10|11) \le L,
\label{chshineq}
\end{align}
where $L=3$ is the local bound, which can be achieved by suitable local deterministic strategies. Such an appropriate strategy is the following. Alice outputs $a=1$ for $x=0$ and $x=1$, and Bob outputs $b=1$ for $y=0$ and $y=1$. That is, the correlations are 
\begin{equation}
P_L(a,b|x,y)=\delta_{a,1}\delta_{b,1},
\label{PL11}
\end{equation}
where 
$\delta_{i,j}$ is the Kronecker delta function: 
\begin{equation}
\delta_{i,j}=
\begin{cases}
1, & \text{if}\ i=j \\
0, & \text{otherwise.}
\end{cases}
\end{equation}

Note that the above form~(\ref{chshineq}) is less common than the standard correlation form of the CHSH inequality~(\ref{chshcorr}), but they are equivalent up to relabelling of the measurement outcomes. Substituting the value of (\ref{Pchsh}) into (\ref{chshineq}) gives the quantum value $Q = (2+\sqrt 2)$ for the CHSH expression~(\ref{chshineq}). This value gives the maximum violation of the CHSH inequality~(\ref{chshineq}), as shown by Tsirelson~\cite{cirel80}. However, this value can be obtained in the ideal case when Alice and Bob's detectors are perfect, that is, the efficiency of the detectors is unity ($\eta_A=\eta_B=1$). We then consider the case of finite efficiency, especially focusing on two limiting cases, $\eta_{sym}=\eta_A=\eta_B$ and $\eta_{asym}=\eta_A, \eta_B=1$. In case of non-detection let Alice and Bob agree to output the value corresponding to the deterministic strategy above (\ref{PL11}), which gives the local bound 3. We then distinguish four cases according to the detection and non-detection events of Alice's and Bob's detectors. Below $M_A$ ($M_B$) denotes the Bell value in the case where only Alice's (Bob's) detectors fire and $X$ denotes the Bell value in the case where none of the detectors fire:
\begin{enumerate}
\item Both Alice's and Bob's detectors fire, which happens with probability $\eta_A\eta_B$, in which case $CHSH = Q = 2+\sqrt 2$. 
\item Only Alice's detectors fire, which happens with probability $\eta_A(1-\eta_B)$, in which case the correlations are $P(a,b|x,y)=(1/2)\delta_{b,1}$ entailing $CHSH = M_A = 2$. In that case Bob's detectors output $b=1$ for every $y=0,1$.
\item Only Bob's detectors fire. This happens with probability $(1-\eta_A)\eta_B$, and we have $P(a,b|x,y)=\delta_{a,1}(1/2)$ resulting in $CHSH = M_B = 2$. In that case, Alice's detector outputs $a=1$ for every $x=0,1$ in case of non-detection.
\item Neither detector fires, which happens with probability $(1-\eta_A)(1-\eta_B)$. In this case, the statistics~(\ref{PL11}) gives the local bound  $L=3$, that is $CHSH = X = 3$. 
\end{enumerate}  

We then obtain the following Bell inequality, which depends on the detection efficiencies:
\begin{align}
I(\eta_A,\eta_B)=&\eta_A(1-\eta_B)M_A + (1-\eta_A)\eta_BM_B  \nonumber\\
&+\eta_A\eta_BQ +(1-\eta_A)(1-\eta_B)X \le L.
\label{etaineq}
\end{align}
 Whenever this inequality is violated, the original Bell inequality is violated with detection efficiencies $\eta_A$ and $\eta_B$. In the case that $X=L$ and from (\ref{etaineq}), we obtain the threshold efficiency for the symmetric case:
\begin{equation}
\eta_{sym} = \frac{2L-M_A-M_B}{Q+L-M_A-M_B}.
\label{eq_sym}
\end{equation}
On the other hand, the following threshold is obtained in the asymmetric case:
\begin{equation}
\eta_{asym} = \frac{L-M_A}{Q-M_A}.                   
\label{eq_asym}
\end{equation}
Note that the latter inequality does not depend on $X$. For the standard single-copy CHSH case, we have the parameters $Q = 2+\sqrt 2$, $M_A = M_B = 2$, and $L = 3$, which, substituted into (\ref{eq_sym}) and (\ref{eq_asym}), give the following values
\begin{align}
\eta_{sym} &= 2(\sqrt 2-1),\nonumber\\
\eta_{asym} &= 1/\sqrt 2, 
\label{etachsh}
\end{align}
reproducing the well-known thresholds~\cite{mermin1986,garg87}.

{\it Correlation-type Bell inequalities.}---
The CHSH inequality discussed above is a special type of correlation inequality. Indeed, if the two-party correlations are defined as  
\begin{equation}
\label{Exy}
E_{x,y}=P(00|xy)+P(11|xy)-P(01|xy)-P(10|xy),
\end{equation}
then (\ref{chshineq}) can be written as follows
\begin{equation}
\rm{\overline{CHSH}}=E_{0,0}+E_{0,1}+E_{1,0}-E_{1,1}\le 2, 
\label{chshcorr}
\end{equation}
where $2$ is the local bound. Consider now generic correlation-type Bell inequalities, in which case the Bell inequality can be expressed as follows 
\begin{equation}
I=\sum_{x=1}^m\sum_{y=1}^m M_{x,y}E_{x,y}\le L,
\label{Icorr}
\end{equation}
where $M_{x,y}$ are the Bell coefficients, $m$ is the number of settings per party and $L$ is the local bound.

Let us show that we can slightly beat the value of $\eta_{sym}$ in (\ref{etachsh}) if $m>2$ settings are available.  For the two-qubit maximally entangled state with traceless observables in Eq.~(\ref{etaineq}), we have $M_A=M_B=0$ . Hence, for a correlation-type Bell inequality~(\ref{Icorr}) with the two-qubit maximum quantum value $Q$ and local bound $L$, we obtain the following thresholds
\begin{align}
\eta_{sym}&=\frac{2}{(Q/L)+1},\nonumber\\
\eta_{asym}&=\frac{L}{Q}
\end{align}
using formulas~(\ref{eq_sym}) and (\ref{eq_asym}).
From the relation of the maximum quantum violation of correlation-type Bell inequalities with two-qubit states and the Grothendieck constant of order three, $K_G(3)$, ~\cite{Grothendieck,Acin2006}, we obtain 
\begin{equation}
1.4359\le \frac{Q}{L}\le 1.4644,
\label{KG3bounds}
\end{equation}
where the upper bound is from Ref.~\cite{Hirsch17}, and the lower bound is from Ref.~\cite{Peter17}. The left-hand side of (\ref{KG3bounds}) gives $\eta_{sym}\le0.8211$ and $\eta_{asym}\le0.6964$. Note that $\eta_{sym}$ is slightly lower than $0.8214$, which is the lowest reported threshold value that can be achieved using the maximally entangled two-qubit state. However, the inequality $A_5$ providing this value is not a correlation-type Bell inequality. 

\section{Multiple copies of the CHSH expression}\label{sec:multichsh}

In this section, we investigate the detection efficiency thresholds for the iterated version of the CHSH inequality~\cite{barrett2002,cleve08}. For two copies we have the double-CHSH expression
\begin{align}
&\text{CHSH}_2(a_1,a_2,b_1,b_2,x_1,x_2,y_1,y_2)\nonumber\\
&=\text{CHSH}(a_1,b_1,x_1,y_1)\times\text{CHSH}(a_2,b_2,x_2,y_2),
\label{CHSH2}
\end{align}
where $\text{CHSH}$ is defined by the expression~(\ref{chshineq}). 
Similarly, the $n$-th iterated version is defined by the following product:
\begin{align}
&\text{CHSH}_n(a_1,\ldots,a_n,b_1,\ldots,b_n,x_1,\ldots,x_n,y_1,\ldots,y_n)\nonumber\\
&=\Pi_{i=1}^n{\text{CHSH}(a_i,b_i,x_i,y_i)},
\label{CHSHn}
\end{align}
where $a_i$ ($b_j$) corresponds to Alice's (Bob's) output for the $i$-th copy ($j$-th copy). Furthermore, $x_i$ ($y_j$) corresponds to Alice's (Bob's) input for the $i$-th copy ($j$-th copy). Therefore, the corresponding $\text{CHSH}_n$ inequality has $m=2^n$ inputs and $o=2^n$ outputs.

\begin{figure}[t!]
\includegraphics[trim=80 0 0 0,clip,width=10cm]{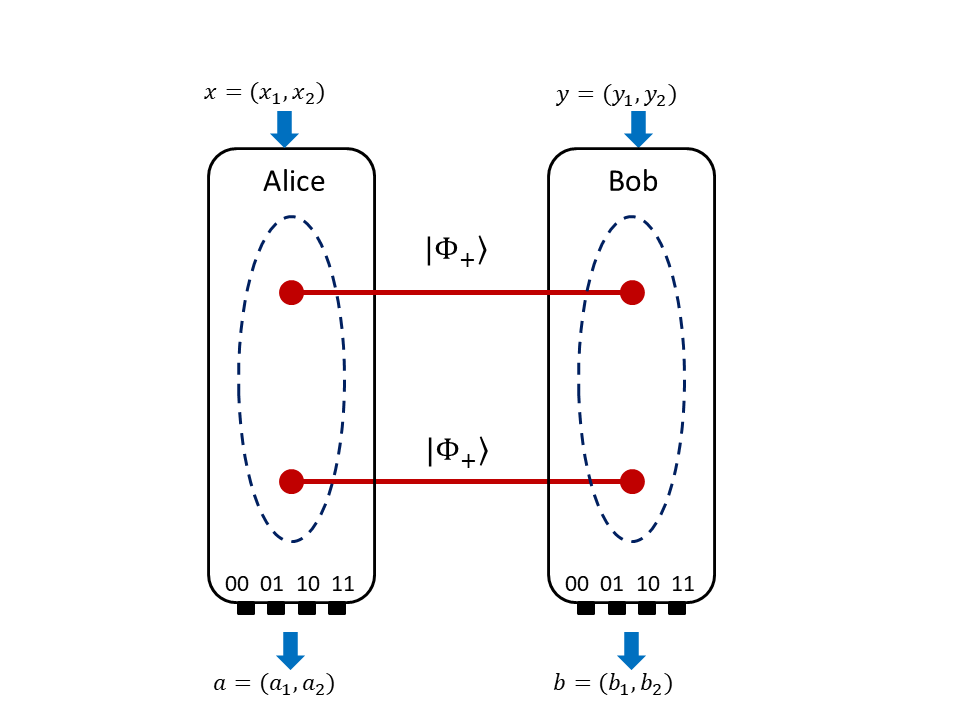}
\caption{Setup for two copies of the CHSH boxes ($n=2$). Here $a_1$ and $a_2$ ($b_1$ and $b_2$) correspond to Alice's (Bob's) output. Similarly, $x_1$ and $x_2$ ($y_1$ and $y_2$) correspond to Alice's (Bob's) inputs. All inputs and outputs are binary, so the total number of inputs and outputs is $2^n$. For $n=2$, this corresponds to a Bell scenario with four inputs and four outputs per party. The quantum maximum is obtained by encoding two copies of the maximally entangled two-qubit states $\ket{\Phi_+}$ into a single pair of particles, and each local measurement (represented by a dashed ellipse) corresponds to a joint two-qubit measurement. The non-detection outcomes are associated with outputs $a=(1,1)$ and $b=(1,1)$.} \label{fig:boxes}
\end{figure}

{\it Quantum value.}---Now let us look at the quantum value of the $\text{CHSH}_n$ expression. We examine the setup with $n$ copies of the two-qubit maximally entangled state and local measurements that are the product of Pauli measurements acting on the $n$-qubit states (see Fig.~\ref{fig:boxes} in the case of $n=2$, where the blue ellipses represent the local measurements). In this case, the probabilities factorize. For example, for the double-CHSH scenario~(\ref{CHSH2}), i.e. $n=2$, we get the following joint correlations 
\begin{align}
&P(a_1,a_2,b_1,b_2|x_1,x_2,y_1,y_2)\nonumber\\
&=P(a_1,b_1|x_1,y_1)\times P(a_2,b_2|x_2,y_2),
\label{P2}
\end{align} 
where the distribution $P(a_i,b_i|x_i,y_i)$ is given by (\ref{Pchsh}). For $n$ copies we have the quantum correlations:
\begin{align}
&P(a_1,\ldots,a_n,b_1,\ldots,b_n|x_1,\ldots,x_n,y_1,\ldots,y_n)\nonumber\\
&=\Pi_{i=1}^n{P(a_i,b_i|x_i,y_i)}.
\label{Pn}
\end{align}
This scenario has $m=2^n$ inputs and $o=2^n$ outputs.

Now we calculate the quantum value $Q^{(n)}$ of the $n$-th iterated CHSH expression $\text{CHSH}_n$. Denote the local bound by $L^{(n)}$. First, we consider the double-CHSH inequality ($n=2$). $\text{CHSH}_2$ is a product of two CHSH expressions, and since the probabilities factorize (see Eq.~(\ref{P2})), we obtain 
\begin{equation}
Q^{(2)}=Q(\text{CHSH})\times Q(\text{CHSH})=Q^2=(2+\sqrt 2)^2.
\label{Qn2} 
\end{equation}
As we see, the quantum value is simply squared. Similarly, for $n$ copies, we have $Q^{(n)}=(2+\sqrt 2)^n$. It is also known~\cite{cleve08} that this value is the Tsirelson bound of the $\text{CHSH}_n$ expression, that is, the maximum quantum value that can be obtained in the presence of arbitrary quantum resources (and in particular when probabilities do not factorize). 

{\it Local bound.}---When calculating the local bound, however, the probabilities do not necessarily factorize with respect to each copy as in Eq.~(\ref{Pn}). For $n=2$ they are as follows:
\begin{align}
&P_L(a_1,a_2,b_1,b_2|x_1,x_2,y_1,y_2)\nonumber\\
=&\sum_{\lambda}P_A(a_1,a_2|x_1,x_2,\lambda)P_B(b_1,b_2|y_1,y_2,\lambda)q(\lambda),
\label{lhv2}
\end{align}
where $\lambda$ is a shared random variable with $\sum_{\lambda}q(\lambda)=1$, and $P_A$ and $P_B$ are arbitrary conditional probability functions (labeled by $\lambda$) on Alice's and Bob's side. Therefore, it is generally allowed to exploit (classical) strategies between different copies. Such a joint strategy is when $a_1$ on Alice's side may depend not only on $x_1$ but also on $x_2$. Indeed, it turns out that the local value $L^{(2)}$ of the Bell expression $\text{CHSH}_2$ is $L^{(2)}=10$, which is larger than $L^2=3^2$. To attain the value of 10, the parties can use the following local deterministic strategies $P_A(a_1,a_2|x_1,x_2)$ and $P_B(b_1,b_2|y_1,y_2)$ in Eq.~(\ref{lhv2}): 
\begin{align}
&P_A(11|00)=P_A(11|01)=P_A(11|10)=P_A(10|11)=1,\nonumber\\
&P_B(11|00)=P_B(11|01)=P_B(11|10)=P_B(10|11)=1
\end{align}
for the respective functions of Alice and Bob. The maximum local value $L^{(2)}=10$ corresponding to the above strategy has been obtained independently by J. Barrett et al.~\cite{barrett2002} and S. Aaronson (see the footnotes in Ref.~\cite{cleve08}). Similarly, for the three-copy $\text{CHSH}_3$ case, we obtain $L^{(3)}=31$. Note that this value is larger than $L(\text{CHSH})\times L(\text{CHSH}_2)=30$. The value of $L^{(3)}=31$ is due to S.~Aaronson and B.~Toner by means of an exhaustive computer search, which was noted in the footnotes of Ref.~\cite{cleve08}. For $n>3$, empirical values of $L^{(4)}=100$, $L^{(5)}=310$ and $L^{(6)}=1000$ are available, which were recently found in Ref.~\cite{araujo2020}. Furthermore, the following analytic upper bound holds for $L^{(n)}$: 
\begin{equation}
L^{(n)}\le (1+\sqrt 5)^n,
\label{Ln}
\end{equation} 
which asymptotically becomes an equality for large $n$. The upper bound~(\ref{Ln}) is due to A.~Ambainis (see Ref.~\cite{yuen}), who builds on Ref.~\cite{dinur2014}. 

{\it Detection efficiencies.}---Let $\eta_{sym}^{(n)}$ and $\eta_{asym}^{(n)}$ denote the symmetric and asymmetric detection efficiency thresholds obtained by $n$-copies of the maximally entangled two-qubit state and anticommuting Pauli measurements. In particular, the $n$-copy distribution~(\ref{Pn}) corresponds to this scenario. We now upperbound the thresholds of $\eta_{sym}^{(n)}$ and $\eta_{asym}^{(n)}$. To this end, we consider $n$ copies of the CHSH expression, that is, the $\text{CHSH}_n$ expression. Any upper bound on the detection efficiency threshold of the $\text{CHSH}_n$ inequality provides an upper bound on the detection efficiencies $\eta_{sym}^{(n)}$ and $\eta_{asym}^{(n)}$ for the case of generic Bell inequalities as well. 

The derivation of the detection-efficiency-dependent Bell inequalities follows the standard procedure. However, see e.g.~Refs.~\cite{massar02,cope19a} for a different way to treat the finite efficiency of the detectors. To take into account inconclusive events, the parties for each of their settings assign one of the valid outcomes to the non-detection event.
This approach has already been discussed in Sec.~\ref{sec:belldet} using the CHSH inequality as an example. Similarly to that case, we associate the non-detection outcome with the particular outcome for which the local deterministic strategy gives the maximum local value of the Bell inequality. We first discuss the $\text{CHSH}_2$ case described by (\ref{CHSH2}), which we later generalize for larger $n$. This setup is depicted in Fig.~\ref{fig:boxes}. 

We now calculate the relevant quantities for this scenario. The quantum value $Q^{(2)}=Q^2=(2+\sqrt 2)^2$ is due to Eq.~(\ref{Qn2}). We will show that $M_A^{(2)}=2^2$ ($M_B^{(2)}=2^2$), where only Alice's (Bob's) detectors fire, respectively.

{\it Proof.}---If only Alice's detectors fire, we have 
\begin{equation}
P(a_1,a_2,b_1,b_2|x_1,x_2,y_1,y_2)=(1/4)P_B(b_1,b_2|y_1,y_2),
\end{equation}
that is, the probability distribution does not depend on Alice's outputs $a=(a_1,a_2)$. Using this probability distribution we get 
\begin{equation}
M_A^{(2)}=\sum_{b_1,b_2,y_1,y_2}P_B(b_1,b_2|y_1,y_2)=4
\end{equation}
for $\text{CHSH}_2$ in (\ref{CHSH2}). This value of 4 is due to two features of the double-CHSH expression: (i) all nonzero Bell coefficients are 1, and (ii) for every $x=(x_1,x_2)$ input of Alice, every $y=(y_1,y_2)$ input of Bob and every output $a=(a_1,a_2)$ of Alice, there is a single nonzero coefficient for Bob's output. That is, the Bell inequality corresponds to a so-called unique game~\cite{khot02}. Note that this value does not depend on the actual deterministic strategy to be used in the case of a non-detection event. Similarly, we obtain $M_B^{(2)}=2^2$ in the case where Bob's detectors fire. $\qed$ 

For general $n$, relying on the two features (i) and (ii) above, we obtain $M_A^{(n)}=M_B^{(n)}=2^n$.    

\begin{figure}[t!]
\includegraphics[trim=0 0 0 0,clip,width=8cm]{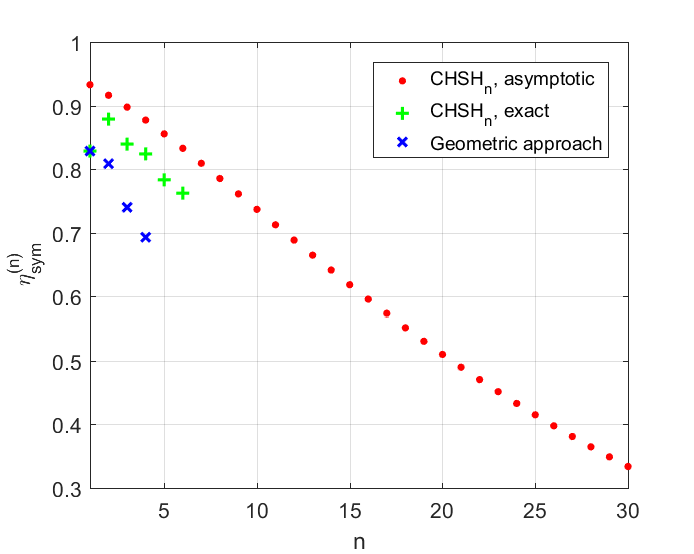}
\caption{Three curves are shown for the upper bounds of the detection efficiency $\eta_{sym}^{(n)}$. The red dots and green crosses are based on the iterated $\text{CHSH}_n$ inequalities. The red curve (dots) is calculated from the formula~(\ref{etasymupper}), while the green curve ($+$ markers) is obtained from the exact $L^{(n)}$ values for $n=1,2,3$. On the other hand, we used their empirical values for $n=4,5,6$, which nevertheless gives very strong evidence of optimality. The values corresponding to the blue $\times$ markers are based on the geometric approach, where the optimal Bell inequality was used.} \label{fig:etasym}
\end{figure}

Plugging the above numbers into (\ref{eq_sym}) and (\ref{eq_asym}), we get upper bounds on the detection efficiencies $\eta_{sym}^{(n)}$ and $\eta_{asym}^{(n)}$ required to see Bell nonlocality. In particular, we have $Q^{(n)}=(2+\sqrt 2)^n$, $M_A^{(n)}=M_B^{(n)}=2^n$ and for $L^{(n)}$ we used the upper bound value in Eq.~(\ref{Ln}). Then we obtain the following upper bounds for $n\ge 1$
\begin{align}
\eta_{\rm sym}^{(n)}&\le\frac{2L^{(n)} - M_A^{(n)}-M_B^{(n)}}{Q^{(n)}+L^{(n)} - M_A^{(n)} -M_B^{(n)}}\nonumber\\
&\le \frac{2((1 + \sqrt{5})^n - 2^n)}{-2^{(n + 1)} + (2 + \sqrt{2})^n + (1 + \sqrt{5})^n}
\label{etasymupper}
\end{align}
and
\begin{equation}
\eta_{asym}^{(n)}\le \frac{L^{(n)}-M_A^{(n)}}{Q^{(n)}-M_A^{(n)}}\le\frac{2^n - (1 + \sqrt{5})^n}{2^n - (2 + \sqrt{2})^n}.           
\label{etaasymupper}   
\end{equation}
We show that it is valid to use the upper bound on $L^{(n)}$ to achieve these bounds. Indeed, we have $Q^{(n)}>L^{(n)}$, $L^{(n)}>M_A^{(n)}$ and $L^{(n)}>M_B^{(n)}$. Then, both (\ref{eq_sym}) and (\ref{eq_asym}) increase when $L^{(n)}$ is replaced by the upper bound~(\ref{Ln}), leading to an upper bound on the detection efficiency thresholds. We note that for $n=1,2,3$, we know the exact $L^{(n)}$ values for $\text{CHSH}_n$, which can be used to give improved upper bounds on $\eta_{\rm sym}^{(n)}$ and $\eta_{asym}^{(n)}$. The corresponding upper bounds for some $n$ values are given in Table~\ref{tab1}. Note that the values given in parentheses are calculated from the empirical values of the local bounds $L^{(n)}$. In these cases, however, we used the see-saw iteration~\cite{Peter17,araujo2020}, which is very efficient for such complexity of problems, and the large number of runs provides very strong evidence of optimality.
Notably, $\eta_{sym}^{(13)}$ beats the $2/3$ limit of Eberhard and $\eta_{asym}^{13}$ beats the $1/2$ limit corresponding to the single-copy CHSH inequality in the case of partially entangled states. Note that, asymptotically, both upper bounds (on $\eta_{sym}^{(n)}$ and $\eta_{asym}^{(n)}$) tend exponentially to zero in the number of copies $n$ (but not in the number of measurement settings $m$ and outcomes $o$, where $m=o=2^n$). The results are shown in Fig.~\ref{fig:etasym}, where the red curve (dots) represents the values given by (\ref{etasymupper}). 

\begin{table}[t!]
	\begin{center}
		\begin{tabular}{|r|c|c|}
			\hline
			$n$ & $\eta_{\rm sym}^{(n)}\le$ & $\eta_{\rm asym}^{(n)}\le$\\
			\hline
			\hline
			$1$  &    $0.8284$&      $0.7071$\\
			\hline
			$2$  &    $0.8787$&      $0.7836$\\
			\hline
			$3$  &    $0.8394$ &     $0.7233$\\
			\hline
			$4$  &    $0.8772 (0.8240)$ &     $0.7813 (0.7007)$\\
			\hline
			$5$  &     $0.8555 (0.7832)$ &     $0.7475 (0.6436)$\\
			\hline
			$6$  &     $0.8328 (0.7622)$ &     $0.7135 (0.6158)$\\
			\hline
			$7$  &     $0.8093$ &     $0.6796$\\
			\hline
			$8$ &     $0.7853$ &     $0.6464$\\
		    \hline
			$9$ &     $0.7610$ &     $0.6142$\\
			\hline
			$10$ &     $0.7367$ &     $0.5832$\\
			\hline
		    $11$ &     $0.7125$ &     $0.5534$\\
			\hline
			$12$ &     $0.7367$ &     $0.6142$\\
			\hline
			$13$ &     $0.6647$ &     $0.4978$\\
			\hline
			$20$ &     $0.5101$ &     $0.3424$\\
			\hline
			$50$ &     $0.1284$ &     $0.0686$\\
			\hline
			$100$&     $0.0094$ &     $0.0047$\\
			\hline
		\end{tabular}
	\end{center}
	\caption{
		Table for the detection efficiency thresholds of the $\text{CHSH}_n$ expression. (First column) The number of copies $n$. Note that the dimension $d$ of the bipartite $(d\times d)$-dimensional system is $2^n$.
		(Second column) The upper bound on the symmetric detection efficiency threshold $\eta_{sym}^{(n)}$ arising from the $\text{CHSH}_n$ expression.
		(Third column) The upper bound on the asymmetric detection efficiency threshold $\eta_{asym}^{(n)}$ arising from the $\text{CHSH}_n$ expression. The values in parentheses are calculated from the empirical value of the local bound of $\text{CHSH}_n$.}
	\label{tab1}
\end{table}

\section{Geometric approach based on linear programming}\label{sec:geom}

We next use the following more general approach to find tighter upper bounds on $\eta_{sym}^{(n)}$ and $\eta_{asym}^{(n)}$. We do not fix the Bell inequality in advance, but we explore the possibly best inequality, where our probability distribution is defined by the $n$-fold product~(\ref{Pn}) of the distribution~(\ref{Pchsh}). 

In this section, we will focus on the case $n=2$ and then in the next section we turn to the cases $n=3,4$. In both cases, we will follow a geometric approach developed by Massar et al.~\cite{massar02}. For $n=2$ we use linear programming, which will provide us with numerically exact values. For higher values of $n$ our geometric approach will be based on Gilbert's algorithm.
Note that in this case, we do not choose a specific Bell inequality in advance. Hence, any upper bound on the detection efficiency threshold based on an optimal Bell inequality obtained in this geometric approach will be at least as low as the threshold calculated for the special $\text{CHSH}_n$ expression.

In order to take into account finite detection efficiency in this method, we modify the probability distribution $P(ab|xy)$. This way we obtain a probability distribution depending on the detection efficiencies. Note that in the previous sections, in contrast the Bell expression itself has depended on the detection efficiencies (see Eq.~(\ref{etaineq}) for the single copy case). To take care of the inconclusive events due to finite detection efficiency, Alice simply chooses the last output $a=(a_1,a_2,\ldots,a_n)=(1,1,\ldots,1)$ for every input $x=(x_1,x_2,\ldots,x_n)$ in case of non-detection. Similarly, for every input $y=(y_1,y_2,\ldots,y_n)$ Bob outputs $b=(b_1,b_2,\ldots,b_n)=(1,1,\ldots,1)$ in case of non-detection. 
As a result, the probabilities apart from the last outcome for Alice and Bob are modified as follows
\begin{align}
P_{\eta_A,\eta_B}(ab|xy) &= \eta_A\eta_B P(ab|xy),\nonumber\\ 
P^A_{\eta_A}(a|x) &= \eta_A P^A(a|x),\nonumber\\
P^B_{\eta_B}(b|y) &= \eta_B P^B(b|y)
\label{Peta}
\end{align}
for all $(x,y)$ and $(a,b)$ except for the outputs $a=(a_1,a_2,\ldots,a_n)=(1,1,\ldots,1)$ and $b=(b_1,b_2,\ldots,b_n)=(1,1,\ldots,1)$. Above $P^A$ and $P^B$ are the marginal distributions of Alice and Bob, defined as follows
\begin{align}
P^A(a|x)=\sum_b P(ab|xy)\, \text{ for all } y,\nonumber\\
P^B(b|y)=\sum_a P(ab|xy)\, \text{ for all } x.
\label{PAPB}
\end{align}
Note that the set of probabilities~(\ref{Peta}) completely determines the probability distribution, since the missing probabilities corresponding to the last $2^n$th outcome of Alice and Bob are completely determined by (\ref{Peta}) due to the no-signaling conditions on the probabilities~(\ref{PAPB}). Hence, from the set of distributions~(\ref{Peta}) we can construct the full set of probabilities:
\begin{equation}
\vec P_{\eta_A,\eta_B}\equiv\{P_{\eta_A,\eta_B}(ab|xy)\}_{a,b,x,y}.
\label{Ppoint}
\end{equation}
Given the above set~(\ref{Ppoint}), our task is to decide whether or not this probability distribution can be described by a local model. For a fixed ($\eta_A$, $\eta_B$) pair of detection efficiencies, this is a feasibility problem, which in turn casts as a linear programming (LP) task. It is noted that there are other ways to handle a non-detection event, see e.g.~Refs.~\cite{massar02,cope19a}. In these works, the non-detection event is treated as an additional outcome for each measurement. For a Bell setup with a given number of settings, the extra outcome may reduce the critical detection efficiency. For instance, in the case of three two-outcome settings per party, the symmetric detection efficiency threshold for the maximally entangled state is reduced from $0.8284$ to $0.8217$ if the non-detection event is treated as an additional outcome~\cite{massar02}. We will see that modeling the detection failure as an additional outcome can also be beneficial in our multi-copy Bell setup. Note, however, that the additional outcome also increases the dimension of the original probability space.

The local set $\mathcal{L}$ for a finite number of inputs $m$ and outputs $o$ is a polytope, the so-called Bell polytope, which is the convex hull of a finite number of points defined by its vertices. The vertices are given by the local deterministic strategies $P_D(ab|xy)=D_A(a|x)D_B(b|y)$, where $D_A(a|x)$ and $D_B(b|y)$ are the deterministic response functions of Alice and Bob, respectively. Alice and Bob each has $o^m$ such functions, so there are $o^{2m}$ deterministic strategies in total, 
\begin{equation}
\vec P_D^{(\lambda)}:=\{P_D^{(\lambda)}(ab|xy)\}_{a,b,x,y},
\label{PD}
\end{equation}
where $\lambda=(1,\ldots,o^{2m})$. Each strategy $\lambda$ translates to a single vertex of the ($o^2m^2$)-dimensional Bell polytope~$\mathcal{L}$. Any point inside this polytope is a convex combination of vertices $\vec P_D^{(\lambda)}$ with some positive weights $q(\lambda)$. 

For the special case of $n$ copies of the probability distribution~(\ref{Pchsh}), we have the Bell scenario $m=o=2^n$. Let us denote the corresponding polytope in this case by $\mathcal{L}^{(n)}$. In particular, if the probability point~$\vec P_{\eta_A,\eta_B}$ lies outside the Bell polytope, it cannot be written as a convex combination of the vertices of the Bell polytope. In this case, we can find the hyperplane separating the polytope $\mathcal{L}^{(n)}$ from the point~$\vec P_{\eta_A,\eta_B}$. This plane is identified with the Bell expression $C$ below the probability point~(\ref{Ppoint}):
\begin{equation}
\sum_{a,b,x,y} C_{a,b,x,y} P_{L}(a,b|x,y)\le 0,
\end{equation}
where $C_{a,b,x,y}$ are the Bell coefficients and $\vec P_L=\{P_L(ab|xy)\}$ is any local distribution satisfying the locality conditions~(\ref{lhv2}). In geometric terms, $\vec P_L$ can be any point located inside the Bell polytope~$\mathcal{L}^{(n)}$. 

To obtain the Bell expression $C$ below the point~(\ref{Ppoint}), we choose fixed parameters $\eta_A$ and $\eta_B$, and solve an LP task as follows: 
\begin{align}
Q\equiv\max_{C_{a,b,x,y}}&  \quad \sum_{a,b,x,y} C_{a,b,x,y} P_{\eta_A,\eta_B}(a,b|x,y)\nonumber\\ 
s.t. &  \sum_{a,b,x,y} C_{a,b,x,y} P_{D}^{(\lambda)}(a,b|x,y)\le 0\, \text{ for all }\lambda,\nonumber\\ 
&  C_{a,b,x,y}\le 1\, \text{ for all } a,b,x,y,
\label{LPtask}
\end{align}
where the index $\lambda$ runs over all local deterministic strategies $\vec P_D^{(\lambda)}$ in Eq.~(\ref{PD}), and the conditions in the last line take care of the upper limit 1 on the coefficients  $C_{a,b,x,y}$. These coefficients are our optimization variables. As mentioned above, there are in total $(o^m)(o^m)$ different local deterministic strategies $\vec P_{D}^{(\lambda)}$. Therefore, in the $n$-copy case, $o=m=2^n$, which amounts to $2^{n 2^{n+1}}$ strategies. In our special case of $n=2$, there are $2^{16}$ strategies. This is the number of vectors we have to provide as an input to the LP, the complexity of which is feasible on a standard desktop computer. 

Note that in the implementation of the algorithm, instead of solving the LP in~(\ref{LPtask}), we have solved a task where the full set of  probabilites~(\ref{Ppoint}) (having dimension $16\times 16$) is replaced by the smaller set~(\ref{Peta}) with dimension $13\times 13$. The corresponding objective function in the optimization~(\ref{LPtask}) is given by
\begin{align}
 &\sum_{a,x}C^A_{a|x}P^A(a|x)+\sum_{b,y}C^B_{b|y}P^B(b|y)\nonumber\\
 &+\sum_{a,b,x,y} C_{a,b,x,y} P_{\eta_A,\eta_B}(a,b|x,y),
\label{LPobj}
\end{align}
where the sum for outputs $a$ and $b$ runs over the first three outputs, that is, $(a_1,a_2)=(0,0),(0,1),(1,0)$ and $(b_1,b_2)=(0,0),(0,1),(1,0)$. However, the sum for inputs $x$ and $y$ runs through all the inputs, that is, $(x_1,x_2)=(0,0),(0,1),(1,0),(1,1)$ for Alice, and similarly for Bob. We used Mosek~\cite{mosek2015} to perform this LP task, which returned the solution to this LP problem within a few seconds.

If the solution to the linear program~(\ref{LPtask}) above is $Q>0$, this indicates that the point~(\ref{Ppoint}) with given $(\eta_A,\eta_B)$ values lies outside the polytope. In the symmetric case, we specify $\eta_{sym}=\eta_A=\eta_B$, and our aim is to choose the smallest $\eta_{sym}$ such that $Q>0$. We do the same for the asymmetric case, $\eta_{asym}=\eta_A$ and $\eta_B=1$. In the limit of the smallest such $\eta_{sym}$ and $\eta_{asym}$, the point~(\ref{Peta}) lies on the boundary of the local set. In the actual computation, $\eta_{sym}$ and $\eta_{asym}$ are chosen such that $Q$ is slightly greater than zero. We next give detailed results for $n=2$ copies of running the above LP problem~(\ref{LPtask}) for both symmetric and asymmetric detection efficiencies. 

{\it Symmetric detection efficiency for two copies.}---Let us focus on the symmetric case for $n=2$. Here we are left with a single parameter, $\eta_{sym}=\eta_A=\eta_B$. Given the  probability distribution $\vec P_{\eta_A,\eta_B}$ in (\ref{LPtask}), our task is to find $\eta_{sym}$ such that the solution $Q$ is some small number (we set $Q$ in the range $0.001$). As a solution to LP~(\ref{LPtask}), we obtain the following form of Bell inequality $I_{sym}\le 0$,
\begin{align}
I_{sym}=&\sum_{a=1}^3\sum_{x=1}^4 C^A_{a|x}P^A(a|x)+\sum_{b=1}^3\sum_{y=1}^4 C^B_{b|y}P^B(b|y)\nonumber\\
&+\sum_{a=1}^3\sum_{b=1}^3\sum_{x=1}^4\sum_{y=1}^4C_{abxy}P(ab|xy)\,, 
\label{Isym}
\end{align}
where Alice's marginal coefficients are $C^A_{1|1}=C^A_{1|2}=C^A_{1|3}=C^A_{1|4}=C^A_{2|2}=C^A_{3|3}=-2$, $C^A_{2|1}=C^A_{3|1}=-1$ and all other entries of $C^A$ are zero. Also, Bob has $C^B=C^A$. On the other hand, the matrix $C$ of size $13\times 13$ is as follows:
\begin{equation}
\resizebox{\columnwidth}{!}{
$\left(\begin{array}{ccc|ccc|ccc|ccc}%
 0  &   0  &   0  &   2  &   1  &  -1  &   2  &  -1  &   1  &   2  &   0  &   0 \\
 0  &   0  &   0  &   1  &   1  &   0  &   0  &   0  &   0  &   1  &   1  &   0 \\
 0  &   0  &   0  &   0  &   0  &   0  &   1  &   0  &   1  &   1  &   0  &   1 \\
\hline
 2 &    1 &    0  &   0  &   1  &  -1  &   2  &   1  &   0  &   0  &   1  &  -1 \\
 1 &    1 &    0  &   1  &   1  &  -1  &   0  &   2  &   0  &   2  &   0  &   0 \\
-1 &    0 &    0  &  -1  &  -1  &  -1  &   1  &   0  &   2  &  -1  &   0  &  -1  \\              
\hline
 2  &   0  &   1  &   2  &   0  &   1  &   0  &  -1   &  1  &   0  &  -1  &   1 \\
-1  &   0  &   0  &   1  &   2  &   0  &  -1  &  -1   & -1  &  -1  &  -1  &   0 \\
 1  &   0  &   1  &   0  &   0  &   2  &   1  &  -1   &  1  &   2  &   0  &   0 \\
\hline
 2  &   1  &   1  &   0  &   2  &  -1  &   0  &  -1  &   2  &  -2  &  -1  &  -1 \\
 0  &   1  &   0  &   1  &   0  &   0  &  -1  &  -1  &   0  &  -1  &  -1  &   0 \\
 0  &   0  &   1  &  -1  &   0  &  -1  &   1  &   0  &   0  &  -1  &   0  &  -1 \\
\end{array}
\right)\,,
$}
\label{Csym}
\end{equation}
where an element $C_{a,b,x,y}$ has been written above as an element $(a,b)$ of the $3\times 3$ submatrix at the coordinate $(x,y)$. Note that a positive multiplicative constant does not change the Bell inequality. In fact, we have doubled the Bell coefficients $C$, $C^A$ and $C^B$ coming from the solution of the LP task to obtain integer values. Also, notice the symmetry of the matrix $C$ with respect to transposition.
 
From this inequality, we can analytically calculate the critical value of $\eta_{sym}$ in the two-copy case, which we denote by $\eta_{sym}^{(2)}$. To this end, we apply formula~(\ref{eq_sym}) to calculate $\eta_{sym}$ given the $Q$, $L$, the $M_A$ ($M_B$) and $X$ values.

We obtain $Q=4(\sqrt{2}-1)$ by substituting (\ref{Csym}) into (\ref{Isym}), where the probabilities are given by the tensor product (\ref{P2}). The local bound $L$ on the other hand is $L=0$. This value can be achieved by a deterministic strategy where for every $x,y$ the fourth output ($a=(1,1)$ and $b=(1,1)$) is given deterministically. Hence $X=L=0$ and the corresponding distribution is $P_L(ab|xy)=\delta_{a,4}\delta_{b,4}$ for every input $(x,y)$. In fact, we have chosen the last outcome for the non-detection event to obtain~(\ref{Peta}). For Alice's non-detection result $P(ab|xy)=(1/4)\delta_{b,4}$, which gives $M_A=-14/4$ for the Bell expression in (\ref{Isym}). Similarly, for Bob's non-detection result, $P(ab|xy)=\delta_{a,4}(1/4)$, which implies $M_B=-14/4$ for the Bell expression in (\ref{Isym}). 
Putting these values together, we have
\begin{equation}
\label{eta0p8086}
\eta_{sym}^{(2)} =\frac{2L-M_A-M_B}{Q+L-M_A-M_B} = \frac{28\sqrt{2}-21}{23}\simeq 0.8086.
\end{equation}
We would like to emphasize that this value is an exact upper bound on $\eta_{sym}^{(2)}$ for the product of Pauli measurements performed on two copies of the maximally entangled two-qubit state. This value is shown in Fig.~\ref{fig:etasym} by the blue cross for $n=2$. 
As mentioned previously, a more general treatment of modeling the detection failure can be achieved by associating an additional outcome with the non-detection event, as opposed to grouping the detection failure with one output. In this more general case, the extra outcome corresponding to non-detection occurs with probability $\eta_A$ on Alice's side and $\eta_B$ on Bob's side, for each measurement independently. In the case of symmetric detection efficiency $\eta_A=\eta_B=\eta$, in the two-copy case ($n=2$), using linear programming we find a threshold of $0.8054$ compared to the threshold of $0.8086$ given by Eq.~(\ref{eta0p8086}). Note that the linear programming approach for $n>2$ is not feasible on a desktop computer. However, we expect from this more general treatment further lowering of the threshold for more than two copies ($n>2$). On the negative side, if such a more general modeling of the failure is used, the dimension of the no-signaling probability space is increased. Recall that for $n=2$, with no additional outcome, the dimension of the no-signaling probability space is $13\times 13$. With one additional outcome per input, however, the no-signaling space becomes $17\times 17$ dimensional.

{\it Asymmetric detection efficiencies for two copies.}---Now consider the asymmetric case for $n=2$, where we have $\eta_{asym}=\eta_A$ and $\eta_B=1$. We solve the LP~(\ref{LPtask}) for $\eta_{asym}$ such that the solution $Q$ is a small number (in the range $0.001$) in the actual computation. As a result, we obtain the Bell inequality $I_{asym}\le 0$ defined similarly to (\ref{Isym}), where 
\begin{align}
I_{asym}=&\sum_{a=1}^3\sum_{x=1}^4 C^A_{a|x}P_A(a|x)+\sum_{b=1}^3\sum_{y=1}^4 C^B_{b|y}P_B(b|y)\nonumber\\
&+\sum_{a,b,x,y}C_{ab|xy}. 
\label{Iasym}
\end{align}
Multiplying all the Bell coefficients by three to get integer values gives
\begin{align}
&C^B_{1|1}=C^B_{1|2}=C^B_{2|2}=C^B_{4|1}=C^B_{4|3}=-1,\nonumber\\
&C^B_{2|1}=C^B_{3|1}=-2 
\label{CB}
\end{align}
and 
\begin{align}
&C^A_{1|1}=C^A_{2|1}=C^A_{3|1}=-3,\nonumber\\
&C^A_{1|2}=C^A_{2|2}=C^A_{1|3}=C^A_{3|3}=-3,\nonumber\\
&C^A_{1|4}=C^A_{2|4}=C^A_{3|4}=-2, 
\label{CA}
\end{align}
and all other coefficients appearing in $C^A$ and $C^B$ are zero. In addition, we have the matrix $C$:
\begin{equation}
\resizebox{\columnwidth}{!}{
$\left(\begin{array}{ccc|ccc|ccc|ccc}
1  &   0  &   0  &   2  &   0  &   0  &   2  &   0  &   0  &   1  &   0  &   0\\
0  &   2  &   0  &   0  &   1  &   0  &   0  &   1  &   0  &   0  &   2  &   0\\
0  &   0  &   2  &   0  &   0  &   1  &   0  &   0  &   1  &   0  &   0  &   2\\
\hline
1  &   0  &   0  &   0  &   2  &   0  &   2  &   0  &   0  &   0  &   1  &   0\\
0  &   2  &   0  &   1  &   0  &   0  &   0  &   1  &   0  &   2  &   0  &   0\\
0  &   0  &   2  &  -1  &  -1  &  -1  &   0  &   0  &   1  &  -2  &  -2  &  -2\\
\hline
1  &   0  &   0  &   2  &   0  &   0  &   0  &   0  &   2  &   0  &   0  &   1\\
0  &   2  &   0  &   0  &   1  &   0  &  -1  &  -1  &  -1  &  -2  &  -2  &  -2\\
0  &   0  &   2  &   0  &   0  &   1  &   1  &   0  &   0  &   2  &   0  &   0\\
\hline
1  &   0  &   0  &   0  &   2  &   0  &   0  &   0  &   2  &  -1  &  -1  &  -1\\
0  &   2  &   0  &   1  &   0  &   0  &  -1  &  -1  &  -1  &   0  &   0  &   2\\
0  &   0  &   2  &  -1  &  -1  &  -1  &   1  &   0  &   0  &   0  &   2  &   0\\
\end{array}
\right)\,,
$}
\label{Casym}
\end{equation}
where an element $C_{abxy}$ is written as an element $(a,b)$ of the $3 \times 3$ submatrix at the coordinate $(x,y)$. Using the inequality~(\ref{Iasym}), we can give analytically the critical value of $\eta_{asym}$ in the two-copy case, denoted by $\eta_{asym}^{(2)}$. We apply formula~(\ref{eq_asym}) for the calculation of $\eta_{asym}$ given the parameters $Q$, $L$, and $M_A$. To this end, we substitute (\ref{Casym}) into (\ref{Iasym}), where the probabilities are given by the tensor product (\ref{P2}), and obtain: 
\begin{equation}
Q=(9/2)(1+\sqrt 2)-9.
\end{equation}
The local bound of (\ref{Casym}) is $L=0$, which can be achieved by a deterministic strategy where for every input ($x,y$) the output is given by the fourth outcome [i.e., $a=b=(1,1)$]. That is, we have the local distribution $P_L(ab|xy)=\delta_{a,4}\delta_{b,4}$ for every $(x,y)$. According to (\ref{Peta}), we have chosen this particular outcome for the non-detection event, and then we get $P(ab|xy)=(1/4)\delta_{b,4}$ when Alice's detector fires, which results in $M_A=-9/4$. Putting these together, we arrive at the following result
\begin{align}
\eta_{asym}^{(2)} =\frac{L-M_A}{Q-M_A} = (1+2\sqrt{2})/7\simeq 0.5469.
\end{align}
This value can be contrasted with the lowest known critical value $\eta_{asym}=0.6520$ among the four-setting two-outcome Bell inequalities~\cite{pal09}. This value in particular corresponds to the $A_{44}$ inequality from the list of Bell inequalities in Ref.~\cite{avis05}.

\section{Geometric approach based on Gilbert's algorithm}\label{sec:geomgilbert}

Unfortunately, the LP~(\ref{LPtask}) used in the preceding section for $n>2$ is not feasible on a standard desktop computer. This is mainly due to the very large number of vectors corresponding to the different deterministic strategies that must be given as an input to the LP problem. Note that for $n=3$ we already have $2^{48}$ different strategies, where each strategy translates to a vector with $4096$ entries.

However, for the $n\ge3$ case we can use an iterative algorithm, the so-called Gilbert algorithm~\cite{gilbert66}, to obtain bounds on $\eta_{sym}^{(n)}$ and $\eta_{asym}^{(n)}$. This algorithm avoids the problem of entering all deterministic strategies in LP and also provides us with the underlying Bell inequality. For $n\le 3$, our method gives correct upper bounds, while for $n=4$ the calculated bound partly relies on heuristic numerical computations. However, we are confident in the validity of the obtained bounds in this case as well.

The values $\eta_{sym}^{(n)}$ and $\eta_{asym}^{(n)}$ we obtain in this section for $n=3$ and $n=4$ are considerably lower than the thresholds corresponding to the case $n=2$ in section~\ref{sec:geom}, and also much lower than the values obtained from the iterated $\text{CHSH}_n$ inequalities for $n=3,4$. We conjecture that the obtained values for $n=3$ and $n=4$ are close to those that could have been obtained by linear programming (assuming the computations could be performed). Here we also give the Bell matrices $C$ obtained by Gilbert's distance method, which are provided as auxiliary data files due to their large size.  

First, we briefly describe Gilbert's distance algorithm~\cite{gilbert66} which is a popular numerical method for collision detection problems (i.e., it detects collisions between rigid convex bodies). In particular, this algorithm estimates the distance between a point $\vec P$ and an arbitrary convex set $\mathcal{S}$ in a finite-dimensional Euclidean space $\RR^d$ via calls to an oracle that performs linear optimizations over the set $\mathcal{S}$. The running time and the convergence properties of the algorithm are very favorable. These properties have been analyzed in detail in Ref.~\cite{brierley16} along with a number of applications in quantum information. A similar method has been used recently in Ref.~\cite{montina2019} to discriminate nonlocal correlations, and further applications in entanglement detection have appeared in Refs.~\cite{shang18,pandya20}.
 
In our particular case, the point $\vec P_{\eta_A,\eta_B}$ is defined by the probability distribution~(\ref{Ppoint}) for the given values of $\eta_A$ and $\eta_B$. Let us first focus on the symmetric case $\eta=\eta_{A}=\eta_{B}$, in which case we obtain a one-parameter family of points $\vec P(\eta)$. We fix $\eta$ such that $\vec P(\eta)$ is outside the local Bell polytope $\mathcal{L}^{(n)}$ (we can take $\eta$ as the best upper bound so far on $\eta_{sym}^{(n)}$). The vertices of the Bell polytope are defined by the deterministic vectors~$\vec P_D^{(\lambda)}$ in (\ref{PD}). For $n$ copies, we have $m=o=2^n$, and there are  $N=o^{2m}=2^{n 2^{n+1}}$ corners of this polytope in dimension $D=(om)^2=2^{4n}$. 

We run Gilbert's distance algorithm, where the inputs to the problem for fixed $\eta$ are the target point $\vec P(\eta)$ and the vertex description $\vec P_D^{(\lambda)}$, $\lambda=(1,\ldots,N)$ of the Bell local polytope $\mathcal{L}^{(n)}$. It is important to keep in mind, however, that this algorithm does not require storing all this data in the computer memory, unlike the linear programming algorithm discussed in Sec.~\ref{sec:geom}. This is a big advantage of the Gilbert method over the LP-based method, since we have already seen that for $n=3$ copies, the number of vertices $N=2^{n 2^{n+1}}=2^{48}$ is too large to be stored in the computer memory. 

Gilbert's algorithm outputs (an estimate to) the distance between the point $\vec P(\eta)$ and the polytope $\mathcal{L}^{(n)}$ by providing a separating hyperplane with normal vector $\vec C$ between the point $\vec P(\eta)$ and the polytope $\mathcal{L}^{(n)}$. We identify this hyperplane with the matrix $C$ of Bell coefficients we are looking for. The description of Gilbert's algorithm adapted to our particular case is given in the Appendix~\ref{sec:app_gib}. In the appendix, we also discuss possible improvements to the algorithm by exploiting symmetry properties of the probability distribution $\vec P(\eta)$.

{\it Detection efficiencies with Gilbert's method for multiple copies.}---
Below we give our computational results on the upper bounds for $\eta_{sym}^{(n)}$ and $\eta_{asym}^{(n)}$ using Gilbert's method discussed above. This includes the $C$ matrices obtained for $n=2,3$ and $4$ copies for both symmetric and asymmetric detection efficiencies. 

We used MATLAB for all the calculations in this paper. The routines \verb|test_sym_n.m| and \verb|test_asym_n.m| test certain properties of the Bell matrices $C$. In the different scenarios, the Bell matrices are named \verb|Csym_n.txt| and \verb|Casym_n.txt|, where \verb|n| denotes the number of copies $n=2,3,4$ and \verb|sym/asym| denotes the case of symmetric/asymmetric detection efficiency. These MATLAB routines and data files are provided as an ancillary file in Ref.~\cite{anc_arxiv}.  

The routines \verb|test_sym_n.m| and \verb|test_asym_n.m| define $n$ copies of the quantum state (\ref{singlet}) and the measurement operators~(\ref{anticomm}), which are used to build up the $n$-copy statistics~(\ref{Pchsh}). From this, the routines compute the following quantities that appear in the formulas~(\ref{eq_sym},\ref{eq_asym}): $Q$, $M_A$, $M_B$, $L$, and $X$. These values are evaluated for the Bell expression $C$. In the computation of $X$, the last outcome is given in the case of a non-detection event.

The values listed above give $\eta_{asym}^{(n)}$ according to the formula (\ref{eq_asym}). On the other hand, the value $\eta_{sym}^{(n)}$ is obtained solving for $\eta$ the quadratic equation 
\begin{equation}
\eta^2 Q + \eta(1-\eta)(M_A+M_B) (1-\eta)^2 X = L.
\label{etaineqX}
\end{equation}

\begin{table}[ht]
	\begin{center}
		\begin{tabular}{|l|c|c|c|}
			\hline
			$ $ & $n=2$ & $n=3$ & $n=4$\\
			\hline
			\hline
			$Q$ &   $7411.71$    &    $2562.88$&      $88170.56$\\
			\hline
			$M_A$ &   $1697.25$    &    $9524.25$&      $35297$\\
			\hline
			$M_B$ &   $1697.25$    &    $9524.25$ &     $35297$\\
			\hline
			$X$ &   $5579$  &     $18949$ &     $60869$\\
			\hline
			$L$ &   $5580$    &    $18979$ &     $63096$\\
			\hline
			$\eta_{sym}^{(n)}$ &  $\le0.8091$  &     $\le0.7399$ &     $\le0.6929$\\
			\hline
		\end{tabular}
	\end{center}
	\caption{Table for calculating upper bounds on the symmetric detection efficiency thresholds $\eta_{sym}^{(n)}=\eta_A =\eta_B$ for $n=2,3$ and $4$ copies. The corresponding Bell expression $C$ has $m=2^n$ inputs and $o=2^n$ outputs. We note that $C$ is invariant under the exchange of parties, so $M_A$ and $M_B$ are equal. The values of $X$ and $L$ are integers, since the Bell coefficients are rounded to integers. The value of $L$ for $n=4$ is a lower bound coming from a heuristic, but is assumed to be exact. Hence, the upper bound to $\eta_{sym}^{(4)}$ is also a conjectured value.}
	\label{tab2sym}
\end{table}

For the case of symmetric detection efficiency, we obtain Table~\ref{tab2sym}. Let us remark that for $n=2$, the result $\eta_{sym}^{(2)}\le 0.8091$ is consistent with the exact $\eta_{sym}^{(2)}=0.8086$ obtained with the LP-based algorithm in Sec.~\ref{sec:geom}. On the other hand, for $n=4$, we had to resort to a heuristic numerical search to obtain the value of $L$, so the obtained value is only a lower bound on $L$. Nevertheless, we still have good confidence in the value due to the efficient numerical procedure used (see \cite{liang09,araujo2020}). Let us also mention that, despite the enormous number of different strategies ($2^{128}$), a branch-and-bound type algorithm may still allow to tackle this problem, similarly to the two-party two-outcome problem used in Ref.~\cite{Peter17}. On the other hand, Table~\ref{tab2asym} presents the asymmetric case.

\begin{table}[ht]
	\begin{center}
		\begin{tabular}{|l|c|c|c|}
			\hline
			$ $ & $n=2$ & $n=3$ & $n=4$\\
			\hline
			\hline
			$Q$ &   $7156.47$    &    $28160.33$&      $108734.28$\\
			\hline
			$M_A$ &   $3766.50$    &    $16009.87$&      $55054.62$\\
			\hline
			$L$ &   $5652$    &    $22308$ &     $79664$\\
			\hline
			$\eta_{asym}^{(n)}$ &  $\le0.5562$  &     $\le0.5183$ &     $\le0.4584$\\
			\hline
		\end{tabular}
	\end{center}
	\caption{
		Table for the asymmetric detection efficiency threshold $\eta_{asym}^{(n)}=\eta_A$ and $\eta_B=1$ for $n=2,3$ and $4$ copies. The Bell expression $C$ has $m=2^n$ inputs and $o=2^n$ outputs. Unlike the case of symmetric detection efficiency, this is not invariant under party exchange. We only present $M_A$, which is required to calculate $\eta_{asym}^{(n)}$ according to (\ref{eq_asym}). The values of $X$ and $L$ are integers, since all Bell coefficients are integers. The value of $L$ for $n=4$ is numerically conjectured. Hence the upper bound to $\eta_{asym}^{(4)}$ is also conjectured.}
	\label{tab2asym}
\end{table}

We note that, similarly to the symmetric case, the upper bound of $0.5562$ for the 2-copy ($n=2$) justifies the usage of Gilbert's method. This value is consistent with the numerically exact value of $0.5469$ computed with LP in Sec.~\ref{sec:geom}. Note that the value $\eta_{asym}^{(4)}$ falls below $1/2$ corresponding to the bound $\eta_{asym}$ for Bell experiments with two inputs and an arbitrary number of outputs. In fact, any Bell test with $N$ inputs will not tolerate $\eta_{asym}$ less than $1/N$~\cite{massar03}. Note also the decreasing upper bound on $\eta_{asym}^{(n)}$ as $n$ increases.

\section{Discussion}\label{sec:discuss}

In this paper, we investigated the critical efficiency of detectors for observing Bell nonlocality using multiple copies of the two-qubit maximally entangled state encoded in a single pair of particles, and the product of qubit Pauli observables acting in the corresponding tensor product of qubit subspaces. The above measurements give the Tsirelson bound of the CHSH inequality for each copy of the state. We showed that the symmetric detection efficiency threshold $\simeq82.84\%$ corresponding to the CHSH-Bell test with the two-qubit maximally entangled state can be considerably lowered by using multiple copies of the state. To this end, we first analytically investigated a special Bell inequality, the $n$th iterative version of the CHSH inequality, and found that the detection efficiency threshold of this composite Bell inequality tends to zero as $n$ increases. For small $n$, we construct Bell inequalities based on a geometric approach which for a given $n$ gives even lower critical detection efficiencies. We used linear programming for $n=2$ copies and Gilbert's algorithm for $n=3$ and $n=4$ copies to obtain Bell inequalities that outperform the $n$th iterated CHSH inequality. 

In the symmetric case, using $n=2,3,4$ copies of the maximally entangled two-qubit state, we find the respective upper bounds of $80.86\%$, $73.99\%$ and $69.29\%$ on $\eta_{sym}^{(n)}$. For the asymmetric case (when one party has unit detection efficiency) the upper bounds of $54.69\%$, $51.83\%$ and $45.84\%$ have been obtained on $\eta_{asym}^{(n)}$ using $n=2,3$ and $4$. The number of measurements and number of outcomes per party for $n=2,3$ and $n=4$ copies are 4, 8 and 16, respectively. 

Note that the above values for $n=4$ are in the same range as the emblematic Eberhard thresholds $\eta_{sym}=2/3$ and $\eta_{asym}=1/2$, which correspond to two partially entangled qubits~\cite{eberhard93}. However, in contrast to Eberhard's result, we used multiple maximally entangled Bell pairs. Both cases have their own advantages in terms of possible technological implementation, and we believe that our setup may offer a promising alternative to the Eberhard setup used in the experiments of Refs.~\cite{giustina15,shalm15} to obtain a loophole-free Bell violation. We also note that very recently similar ideas, based in part on our current methods, have been used to obtain bipartite Bell inequalities with very low critical detection efficiency~\cite{Miklin22,Xu22,GPC22}. On a different note, we also mention that in a broadcast scenario~\cite{Bowles21} using a single copy of a two-qubit maximally entangled state, one can achieve the detection efficiency threshold $0.7355$ as recently shown in Ref.~\cite{Boghiu21}.

\section{Acknowledgements} 

We thank Yeong-Cherng Liang, Marco T. Quintino and G\'eza T\'oth for interesting discussions. We acknowledge the support of the EU (QuantERA eDICT) and the National Research, Development and Innovation Office NKFIH (No. 2019-2.1.7-ERA-NET-2020-00003).

\appendix

\section{Gilbert's algorithm adapted to the detection efficiency problem}
\label{sec:app_gib}

In this appendix, we discuss the Gilbert algorithm adapted to the detection efficiency problem in Bell setups. Gilbert's algorithm outputs (an estimate to) the distance between
a point $\vec P$ and an arbitrary convex set $S$ by calling an oracle that carries out a linear optimization over $S$. In our case the point is $\vec P(\eta)$ from (\ref{Ppoint}) (wherein the symmetric case we set $\eta=\eta_{A}=\eta_{B}$) and the convex set is the local Bell polytope $\mathcal{L}^{(n)}$ defined by the vertices in (\ref{PD}). The algorithm provides a separating hyperplane with normal vector $\vec C$ between the point $\vec P(\eta)$ and the polytope $\mathcal{L}^{(n)}$. We identify this hyperplane with the matrix $C$ of Bell coefficients we are looking for. The algorithm in our particular case is defined as follows~\cite{gilbert66,brierley16}: 

\noindent {\it Inputs}: the vector $\vec P(\eta)$ specified by the number of copies $n$ and the parameter $\eta$ and the description of the polytope $\mathcal{L}^{(n)}$. The steps are as follows.
\begin{enumerate}
\item Set $k=0$ and a value of $\epsilon$ (typically small), and pick an arbitrary point $\vec P_k$ within the polytope $\mathcal{L}^{(n)}$. 
\item Given the point $\vec P_k$ and the target point $\vec P(\eta)$, run an oracle, that maximizes the overlap $(\vec P(\eta)-\vec P_k)\cdot\vec P_D^{(\lambda)}$ over all vertices $\vec P_D^{(\lambda)}\in\mathcal{L}$, $\lambda=1,\ldots,N$, where $N=2^{n 2^{n+1}}$. Denote the index of the local deterministic point returned by the oracle by $k'$ and the corresponding point by $\vec P_D^{(k')}$.
\item Let us find the point $\vec P_{k+1}$ as the convex combination of $\vec P_k$ and $\vec P_D^{(k')}$ that minimizes the distance $\|\vec P(\eta) - \vec P_{k+1}\|$.
\item Let $k=k+1$ and go to Step 2 until the distance $\|\vec P(\eta) - \vec P_{k}\|\le\epsilon$. 
\end{enumerate}
\noindent {\it Output}: $\vec C\equiv \vec P(\eta) - \vec P_{k}$.

The Bell matrix $C$ is then identified with the returned solution vector $\vec C$. Below we give possible modifications to the above algorithm.

In step 2 we have to maximize the overlap $(\vec P(\eta)-\vec P_k)\cdot\vec P_D^{(\lambda)}$ over $N=2^{n 2^{n+1}}$ deterministic vectors. This number is exponential in the number of measurement settings $m=2^n$. In fact, this maximization task is an NP-hard problem~\cite{pitowsky89}, and it seems unlikely to find an efficient solution in the general case. Therefore, we resort to a heuristic search instead of the exact enumeration method. The description of this method can be found in Refs.~\cite{brierley16,Hirsch17} for the special case of two outcomes and in Refs.~\cite{liang09,araujo2020} for the case with more than two outcomes. 

The returned vector $\vec C$ has entries $C_{abxy}$, where $a,b,x,y=(1,\ldots,2^n)$. This $\vec C$ corresponds to a separating hyperplane, which separates the point $\vec P(\eta)$ from the Bell polytope $\mathcal{L}^{(n)}$. From $\vec C$ we can produce the matrix $C$, where the element $C_{abxy}$ is written as the element $(a,b)$ of the $2^n \times 2^n$ submatrix at the coordinate $(x,y)$. Note, however, that the oracle in step 2 has a heuristic nature. Therefore, we also run a brute force computation by enumerating all the $2^{n 2^{n+1}}$ strategies to check that the local bound for the Bell expression $C$ is given correctly. This check has been carried out for $n=2$ and $n=3$, but the case of $n=4$ is computationally hard to tackle, so in the latter case, our result is based partly on a heuristic computation.

We can add to step 3 a modification introduced in Ref.~\cite{brierley16}. In this case, when finding a point $\vec P_{k+1}$, we keep not only $\vec P_k$ but also the previous $m$ points $\vec P_{k-1}, \vec P_{k-2},\ldots,\vec P_{k-m}$ and find a convex combination of all of these to minimize the distance to $\vec P(\eta)$. This optimization can be done efficiently for not too large $m$ by solving a linear least squares problem. In our actual computations, we set the value $m$ in the range $m=20,\ldots,100$.

In addition, we can build a symmetrization procedure in step 3. Here, we exploit the fact that the distribution $P(ab|xy)$ in formula~(\ref{Pn}) is invariant under the simultaneous permutation of Alice and Bob devices. Hence in the case of $n=2$ we can simultaneously swap Alice's and Bob's devices without changing the distribution $P(ab|xy)$. For $n$ copies we have $n!$ such permutations between the devices. We impose this symmetry on the Bell functional $C$ as well. Namely, for $n=2$ let us have $C_{a,b,x,y}=\tilde C_{a,b,x,y}$ for all $a,b,x,y$, where we define 
\begin{align}
C_{a,b,x,y}&\equiv C_{a_1,a_2,b_1,b_2,x_1,x_2,y_1,y_2}\nonumber\\
\tilde C_{a,b,x,y}&\equiv C_{a_2,a_1,b_2,b_1,x_2,x_1,y_2,y_1}. 
\end{align}
Then from the vector $\vec P_D^{(k')}$ in step 3 of the above algorithm we form the symmetrized vector
\begin{equation}
\frac{P_D^{(k')}+\tilde P_D^{(k')}}{2}, 
\label{PDper2}
\end{equation} 
where the components of $\tilde P_D^{(k')}$ are given by 
\begin{equation}
\tilde P_D^{(k')}(a,b|x,y)\equiv P_D^{(k')}(a_2,a_1,b_2,b_1|x_2,x_1,y_2,y_1). 
\label{PDswitch}
\end{equation}
Notice that in step 2, the symmetrized vector~(\ref{PDper2}) gives the same overlap with $[\vec P(\eta)-\vec P_k]$ as $\vec P_D^{(k')}$ does. 
On the other hand, at the end of the procedure, we obtain a Bell matrix $C$ with the required symmetry $C=\tilde C$. For $n>2$, the symmetrization is similar to the above procedure. In the general case of $n$ copies, there are $n!$ different possible permutations of the devices, all of which have to be taken into account in the symmetrization task.


\begin{thebibliography}{99}
	
\bibitem{Bell64}
J.~S. Bell.
\newblock {On the Einstein-Poldolsky-Rosen paradox}.
\newblock {\em Physics} {\bf 1}, 195--200 (1964).

\bibitem{FC}
S. Freedman and J. Clauser.
\newblock Experimental test of local hidden-variable theories.
\newblock {\em Phys. Rev. Lett.} {\bf 28}, 938 (1972).

\bibitem{larsson14}
J.-{\AA}. Larsson.
\newblock Loopholes in Bell inequality tests of local realism.
\newblock {\em Journal of Physics A: Mathematical and Theoretical}
{\bf 47}, 424003 (2014).

\bibitem{Aspect}
A. Aspect, J. Dalibard, and G. Roger.
\newblock Experimental test of Bell's inequalities using time-varying
analyzers.
\newblock {\em Phys. Rev. Lett.} {\bf 49}, 1804--1807 (1982).

\bibitem{hensen15}
B.~Hensen, H.~Bernien, A.~E. Dr{\'{e}}au, A.~Reiserer, N.~Kalb, M.~S. Blok,
J.~Ruitenberg, R.~F.~L. Vermeulen, R.~N. Schouten, C.~Abell{\'{a}}n,
W.~Amaya, V.~Pruneri, M.~W. Mitchell, M.~Markham, D.~J. Twitchen, D.~Elkouss,
S.~Wehner, T.~H. Taminiau, and R.~Hanson.
\newblock Loophole-free Bell inequality violation using electron spins
separated by 1.3 kilometres.
\newblock {\em Nature} {\bf 526}, 682--686 (2015).

\bibitem{giustina15}
M. Giustina, M.~A.~M. Versteegh, S. Wengerowsky, J. Handsteiner, A. Hochrainer, K. Phelan, F. Steinlechner, J. Kofler, J.-{\AA}. Larsson, C. Abell{\'a}n, et~al.
\newblock Significant-loophole-free test of Bell's theorem with entangled
photons.
\newblock {\em Phys. Rev. Lett.} {\bf 115}, 250401 (2015).

\bibitem{shalm15}
L.~K. Shalm, E. Meyer-Scott, B.~G. Christensen, P. Bierhorst, M.~A. Wayne, M.~J. Stevens, T. Gerrits, S. Glancy, D.~R. Hamel, M.~S. Allman, et~al.
\newblock Strong loophole-free test of local realism.
\newblock {\em Phys. Rev. Lett.} {\bf 115}, 250402 (2015).

\bibitem{rosenfeld2017}
W. Rosenfeld, D. Burchardt, R. Garthoff, K. Redeker, N. Ortegel, M. Rau, and H. Weinfurter.
\newblock Event-ready Bell test using entangled atoms simultaneously closing
detection and locality loopholes.
\newblock {\em Phys. Rev. Lett.} {\bf 119}, 010402 (2017).

\bibitem{pearle}
P.~M. Pearle.
\newblock Hidden-variable example based upon data rejection.
\newblock {\em Phys. Rev. D} {\bf 2}, 1418 (1970).

\bibitem{sciarrino11}
F. Sciarrino, G. Vallone, A. Cabello, and P. Mataloni.
\newblock Bell experiments with random destination sources.
\newblock {\em Phys. Rev. A} {\bf 83}, 032112 (2011).

\bibitem{kost18}
K. Kostrzewa, W. Laskowski, and T. Vertesi.
\newblock Closing the detection loophole in multipartite Bell experiments with a limited number of efficient detectors.
\newblock {\em Phys. Rev. A} {\bf 98}, 012138 (2018).

\bibitem{cope19a}
T. Cope and R. Colbeck.
\newblock Bell inequalities from no-signaling distributions.
\newblock {\em Phys. Rev. A} {\bf 100}, 022114 (2019).

\bibitem{CHSH}
J.~F. Clauser, M.~A. Horne, A. Shimony, and R.~A. Holt.
\newblock Proposed experiment to test local hidden-variable theories.
\newblock {\em Phys. Rev. Lett.} {\bf 23}, 880--884 (1969).

\bibitem{mermin1986}
N.~D. Mermin.
\newblock The EPR experiment--thoughts about the ``loophole''.
\newblock {\em Annals of the New York Academy of Sciences} {\bf 480}, 422--427 (1986).

\bibitem{garg87}
A. Garg and N.~D. Mermin.
\newblock Detector inefficiencies in the Einstein-Podolsky-Rosen experiment.
\newblock {\em Phys. Rev. D} {\bf 35}, 3831 (1987).

\bibitem{brunner_gisin}
N. Brunner and N. Gisin.
\newblock Partial list of bipartite Bell inequalities with four binary settings.
\newblock {\em Physics Letters A} {\bf 372}, 3162--3167 (2008).

\bibitem{avis05}
D. Avis, H. Imai, T. Ito, and Y. Sasaki.
\newblock Two-party Bell inequalities derived from combinatorics via triangular
elimination.
\newblock {\em Journal of Physics A: Mathematical and General} {\bf 38}, 10971 (2005).

\bibitem{massar02}
S. Massar, S. Pironio, J. Roland, and B. Gisin.
\newblock Bell inequalities resistant to detector inefficiency.
\newblock {\em Phys. Rev. A} {\bf 66}, 052112 (2002).

\bibitem{detasym1}
A. Cabello and J.-{\AA}. Larsson.
\newblock Minimum detection efficiency for a loophole-free atom-photon Bell experiment.
\newblock {\em Phys. Rev. Lett.} {\bf 98}, 220402 (2007).

\bibitem{detasym2}
N. Brunner, N. Gisin, V. Scarani, and C. Simon.
\newblock Detection loophole in asymmetric Bell experiments.
\newblock {\em Phys. Rev. Lett.} {\bf 98}, 220403 (2007).

\bibitem{garbarino10}
G. Garbarino.
\newblock Minimum detection efficiencies for a loophole-free observable-asymmetric Bell-type test.
\newblock {\em Phys. Rev. A} {\bf 81}, 032106 (2010).

\bibitem{gisin99}
N. Gisin and B. Gisin.
\newblock A local hidden variable model of quantum correlation exploiting the
detection loophole.
\newblock {\em Phys. Lett. A} {\bf 260}, 323--327 (1999).

\bibitem{massar02nonlocality}
S. Massar.
\newblock Nonlocality, closing the detection loophole, and communication
complexity.
\newblock {\em Phys. Rev. A} {\bf 65}, 032121 (2002).

\bibitem{eberhard93}
P.~H. Eberhard.
\newblock Background level and counter efficiencies required for a
loophole-free Einstein-Podolsky-Rosen experiment.
\newblock {\em Phys. Rev. A} {\bf 47}, R747 (1993).

\bibitem{hardy93}
L. Hardy.
\newblock Nonlocality for two particles without inequalities for almost all
entangled states.
\newblock {\em Phys. Rev. Lett.} {\bf 71}, 1665 (1993).

\bibitem{gomez19}
S.~G{\'o}mez, A.~Mattar, I.~Machuca, E. S.~G{\'o}mez, D.~Cavalcanti, O.~Jim{\'e}nez
Far{\'\i}as, A.~Ac{\'\i}n, and G.~Lima.
\newblock Experimental investigation of partially entangled states for
device-independent randomness generation and self-testing protocols.
\newblock {\em Phys. Rev. A} {\bf 99}, 032108 (2019).

\bibitem{colbeck09}
R.~{Colbeck}.
\newblock {Quantum and relativistic protocols for secure multi-party computation}.
\newblock {\em e-print arXiv:0911.3814} (2009)

\bibitem{pironio10}
S. Pironio, A. Ac{\'\i}n, S. Massar, A~Boyer de~La~Giroday, D.~N. Matsukevich, P. Maunz, S. Olmschenk, D. Hayes, L.~Luo, T.~A. Manning, and C. Monroe.
\newblock Random numbers certified by Bell's theorem.
\newblock {\em Nature} {\bf 464}, 1021--1024 (2010).

\bibitem{barrett2002}
J. Barrett, D. Collins, L. Hardy, A. Kent, and S. Popescu.
\newblock Quantum nonlocality, Bell inequalities, and the memory loophole.
\newblock {\em Phys. Rev. A} {\bf 66} 042111 (2002).

\bibitem{VPN10}
T. V{\'e}rtesi, S. Pironio, and N. Brunner.
\newblock Closing the detection loophole in Bell experiments using qudits.
\newblock {\em Phys. Rev. Lett.} {\bf 104}, 060401 (2010).

\bibitem{cope19b}
T. Cope.
\newblock The role of entanglement in quantum communication, and analysis of
the detection loophole.
\newblock {\em e-print arXiv:1904.11769} (2019).

\bibitem{kwiat97}
P.~G. Kwiat.
\newblock{Hyper-entangled states}.
\newblock {\em J. Mod. Optics} {\bf 44}, 11--12 (1997).

\bibitem{erhard2020}
M. Erhard, M. Krenn, and A. Zeilinger.
\newblock{Advances in high-dimensional quantum entanglement}.
\newblock {\em Nat. Rev. Phys.} {\bf 2}, 365 (2020).

\bibitem{genovese08}
M. Genovese, P. Traina.
\newblock{Review on qudits production and their application to quantum communication and studies on local realism}.
\newblock {\em Adv. Sci. Lett.} {\bf 1}, 153--160 (2008).

\bibitem{CGLMP02}
D. Collins, N. Gisin, N. Linden, S. Massar, and S. Popescu.
\newblock Bell inequalities for arbitrarily high-dimensional systems.
\newblock {\em Phys. Rev. Lett.} {\bf 88}, 040404 (2002).

\bibitem{cirel80}
B.~S. Cirel'son.
\newblock Quantum generalizations of Bell's inequality.
\newblock {\em Letters in Mathematical Physics} {\bf 4}, 93--100 (1980).

\bibitem{Grothendieck}
A.~Grothendieck.
\newblock R\'esum\'e de la th\'eorie m\'etrique des produits tensoriels
topologiques.
\newblock {\em Bol. Soc. Mat. S\~ao Paulo} {\bf 8}, 1--79 (1953).

\bibitem{Acin2006}
A. Ac\'in, N. Gisin, and B. Toner.
\newblock Grothendieck's constant and local models for noisy entangled
quantum states.
\newblock {\em Phys. Rev. A} {\bf 73}, 062105 (2006).

\bibitem{Hirsch17}
F. Hirsch, M.~T. Quintino, T. V\'ertesi, M. Navascu\'es, and N. Brunner.
\newblock Better local hidden variable models for two-qubit Werner states and an upper bound on the Grothendieck constant $K_G(3)$.
\newblock {\em Quantum} {\bf 1}, 3 (2017).

\bibitem{Peter17}
P. Divi\'anszky, E. Bene, and T. V\'ertesi.
\newblock Qutrit witness from the Grothendieck constant of order four.
\newblock {\em Phys. Rev. A}, {\bf 96}, 012113 (2017).

\bibitem{cleve08}
R. Cleve, W. Slofstra, F. Unger, and S. Upadhyay.
\newblock Perfect parallel repetition theorem for quantum xor proof systems.
\newblock {\em Computational Complexity}, {\bf 17}, 282--299 (2008).

\bibitem{araujo2020}
M. Ara{\'u}jo, F. Hirsch, and M.~T. Quintino.
\newblock Bell nonlocality with a single shot.
\newblock {\em Quantum} {\bf 4}, 353 (2020).

\bibitem{yuen}
H. Yuen, \url{https://www.microsoft.com/en-us/research/wp-content/uploads/2017/09/2017-01-18-Session-VB-Henry-Yuen.pdf}.

\bibitem{dinur2014}
I. Dinur and D. Steurer.
\newblock Analytical approach to parallel repetition.
\newblock In {\em Proceedings of the forty-sixth annual ACM symposium on Theory
	of computing}, pages 624--633, 2014.

\bibitem{khot02}
S. Khot.
\newblock On the power of unique 2-prover 1-round games.
\newblock In {\em Proceedings of the thiry-fourth annual ACM symposium on
	Theory of computing}, pages 767--775, 2002.

\bibitem{mosek2015}
Mosek ApS.
\newblock The MOSEK optimization toolbox for MATLAB manual, 2015.

\bibitem{pal09}
K.~F. P{\'a}l and T. V{\'e}rtesi.
\newblock Quantum bounds on Bell inequalities.
\newblock {\em Physical Review A}, {\bf 79}, 022120 (2009).

\bibitem{gilbert66}
E.~G. Gilbert.
\newblock An iterative procedure for computing the minimum of a quadratic form
on a convex set.
\newblock {\em SIAM Journal on Control} {\bf 4}, 61--80 (1966).

\bibitem{brierley16}
S. Brierley, M. Navascu\'es, and T. V\'ertesi.
\newblock Convex separation from convex optimization for large-scale problems.
\newblock {\em e-print arXiv:1609.05011} (2016).

\bibitem{montina2019}
A. Montina and S. Wolf.
\newblock Discrimination of non-local correlations.
\newblock {\em Entropy} {\bf 21}, 104 (2019).

\bibitem{shang18}
J. Shang and O. G{\"u}hne.
\newblock Convex optimization over classes of multiparticle entanglement.
\newblock {\em Phys. Rev. Lett.} {\bf 120}, 050506 (2018).

\bibitem{pandya20}
P. Pandya, O. Sakarya, and M. Wie{\'s}niak.
\newblock Hilbert-Schmidt distance and entanglement witnessing.
\newblock {\em Phys. Rev. A} {\bf 102}, 012409 (2020).

\bibitem{anc_arxiv}
I. M\'arton, E. Bene, T. V\'ertesi.
\newblock Ancillary files in \url{https://arxiv.org/abs/2103.10413}.

\bibitem{massar03}
S. Massar and S. Pironio.
\newblock Violation of local realism versus detection efficiency.
\newblock {\em Phys. Rev. A} {\bf 68}, 062109 (2003).

\bibitem{Miklin22}
N. Miklin, A. Chaturvedi, M. Bourennane, M. Paw{\l}owski, and A. Cabello.
\newblock Exponentially decreasing critical detection efficiency for any Bell inequality.
\newblock {\em Phys. Rev. Lett.} {\bf 129}, 230403 (2022).

\bibitem{Xu22}
Z. P. Xu, J. Steinberg, J. Singh, A. J. L\'opez-Tarrida, J. R. Portillo, A. Cabello.
\newblock Graph-theoretic approach to Bell experiments with low detection efficiency.
\newblock {\em e-print arXiv:2205.05098} (2022).

\bibitem{GPC22}
J. R. Gonzales-Ureta, A. Predojevi\'c, A. Cabello.
\newblock Optimal and tight Bell inequalities for state-independent contextuality sets.
\newblock {\em e-print arXiv:2207.08850} (2022).

\bibitem{Bowles21}
J. Bowles, F. Hirsch, and D. Cavalcanti.
\newblock Single-copy activation of Bell nonlocality via broadcasting of quantum states.
\newblock {\em Quantum} {\bf 5}, 499 (2021).

\bibitem{Boghiu21}
E. C. Boghiu, F. Hirsch, F., P. S. Lin, M. T. Quintino, J. Bowles. 
\newblock Device-independent and semi-device-independent entanglement certification in broadcast Bell scenarios.
\newblock {\em e-print arXiv:2111.06358} (2021).

\bibitem{pitowsky89}
I. Pitowsky.
\newblock {\em Quantum probability-quantum logic}.
\newblock Springer, New York, 1989.

\bibitem{liang09}
Y.-C. Liang, C.-W. Lim, D.-L. Deng.
\newblock Reexamination of a multisetting Bell inequality for qudits.
\newblock {\em Phys. Rev. A} {\bf 80}, 052116 (2009).

\end{thebibliography}
\end{document}